# Single femtosecond laser pulse-driven ferromagnetic switching

Chen Xiao[1,2,3,7], Boyu Zhang[1,7], Xiangyu Zheng[1,7], Yuxuan Yao[1,7], Jiaqi Wei[1,7], Dinghao Ma[1], Yuting Gong[4], Rui Xu[1], Xueying Zhang[1,2], Yu He[1], Wenlong Cai[1,2], Yan Huang[1], Daoqian Zhu[1,2], Shiyang Lu[1], Kaihua Cao[1], Hongxi Liu[5], Pierre Vallobra[2], Xianyang Lu[4], Youguang Zhang[3], Bert Koopmans[6] and Weisheng Zhao[1,2]*

[1]Fert Beijing Institute, School of Integrated Circuit Science and Engineering, Beihang University, Beijing, 100191, China.

[2]State Key Laboratory of Spintronics, Hangzhou International Innovation Institute, Beihang University, Hangzhou, 311115, China.

[3]School of Electronic Information Engineering, Beihang University, Beijing 100191, China

[4]State Key Laboratory of Spintronics, Nanjing University, Suzhou, 215163, China.

[5]Truth Memory Corporation, Beijing, 100088, China.

[6]Department of Applied Physics, Institute for Photonic Integration, Eindhoven University of Technology, 5600 MB Eindhoven, The Netherlands.

[7]Authors contributed equally to this work.

*Corresponding author: weisheng.zhao@buaa.edu.cn.

**Abstract**

**Light pulses offer a faster, more energy-efficient, and direct route to magnetic bit writing, pointing toward a hybrid memory and computing paradigm based on photon transmission and spin retention. Yet progress remains hindered, as deterministic, single-pulse optical toggle switching has so far been achieved only with ferrimagnetic materials, which require too specific a rare-earth composition and temperature conditions for technological use. In mainstream ferromagnets—central to spintronic memory and storage—such bistable switching is considered fundamentally difficult, as laser-induced heating does not inherently break time-reversal symmetry. Here, we report coherent magnetization switching in ferromagnets, driven by thermal anisotropy torque with single laser pulses. The toggle switching behavior is robust over a broad range of pulse durations, from femtoseconds to picoseconds, a prerequisite for practical applications. Furthermore, the phenomenon exhibits reproducibility in CoFeB/MgO-based magnetic tunnel junctions with a high magnetoresistance exceeding 110%, as well as the scalability down to nanoscales with remarkable energy efficiency (17 fJ per 100-nm-sized bit). These results mark a notable step toward integrating opto-spintronics into next-generation memory and storage technologies.**

**Introduction**

Laser pulses are the fastest way to manipulate magnetization, enabling the bistable transition of magnetic states at femto- and picosecond timescales[1]. Despite the original discovery of sub-picosecond demagnetization in ferromagnetic (FM) nickel[2], single-pulse all-optical helicity-independent switching (AO-HIS) has been predominantly realized in Gd-based ferrimagnetic alloys or multilayers[3–6]. The toggle switching effect is attributed to the mediation of two antiferromagnetically exchange coupled sublattices from rare earth (RE) and transition metal (TM). Although this phenomenon can be realized in synthetic ferrimagnets[7], MnRuGa alloys[8] and exchange biased[9] or coupled composite layers[10–12], it still requires too specific of a composition and temperature range for industrial applications. In comparison, ferromagnetic 3d transition metals, such as Fe and Co possess significantly higher Curie temperatures and superior durability, making them widely adopted in memory and storage industries[13,14]. Therefore, the optical control of interfacial perpendicular magnetic anisotropy (PMA) in ferromagnetic thin films and nanoscale devices such as CoFeB/MgO/CoFeB junctions[15,16] has attracted considerable interest.

Unlike ferrimagnets, ferromagnets intrinsically lack internal dual-sublattice structure necessary for angular momentum exchange during AO-HIS, remaining a fundamental issue for practical applications. As thermal energy alone just reduces magnetization temporarily and restores it randomly or back to the initial state via anisotropy, by itself cannot break the symmetry. To overcome this problem, two approaches have been proposed for single pulse optical switching. The first one uses angular moment in the light (all-optical helicity-dependent switching, AO-HDS)[17,18]; however, the speed of magnetic domain formation requiring multi-pulses is slow. A very recent solution uses minority spins carried by laser-excited hot electrons[19,20] in spin-valve structures[21], whereas the thick metallic spacers may be harder to scale[22].

In this *Article*, we report the demonstration of bistable toggle switching of a single ferromagnetic thin layer, induced by single laser pulses under an in-plane magnetic field. This effect is then elucidated in terms of coherent precessional switching driven by thermal anisotropy torques ($\tau_{\text{TAT}}$)[23,24]. Further experiments also support field-free switching with $\tau_{\text{TAT}}$. With this mechanism, we could extend the pulse durations of optical switching from a few femtoseconds to hundreds of picoseconds. Moreover, ultrafast and energy-efficient reversal also occurs in magnetic tunneling junctions (MTJs), which are rare-earth-free and easy to be read. We emphasize that these findings would lead to innovations in integrating ultrafast magnetism for opto-spintronics.

**Laser-induced manipulation of interfacial perpendicular ferromagnetism**

We use both anomalous Hall effect (AHE) and Kerr effect to probe the magnetic states under laser excitations. The set up is shown in Fig. 1a and methods. Next, a series of thin films X/CoFeB (0.8~1.4)/MgO (2)/Ta (2) (numbers in nm) were prepared on SiO$_2$ substrates with magnetron sputtering, where different bottom buffer layers (X =

Ta, Ti, W, Ta/Ti) were prepared for a systematic investigation. Samples were annealed ensuring a better interfacial PMA except for X = Ti and Ta/Ti, see methods. The thin films were then patterned into cross-shaped Hallbar structures for optical access and Hall voltages readout (Fig. 1a, under). In this way, the AHE resistances can quantify the changes of the perpendicular magnetization $M_z$ induced by laser stimulus assisted by the magnetic field. Figure 1b shows a standard $R$-$H_z$ loop of sample Ti (2)/CoFeB (1)/MgO (2)/Ta (2) measured under a sweeping perpendicular magnetic field $H_z$. We initially perform the measurements in the absence of $H_x$, recording $M_z$ during a circle sweep of $H_z$ (Fig. 1b). To study the effects of an ultrashort light pulse on $M_z$, we then apply ~50-fs-width pulses of variable power supplied by a Ti:Sapphaire laser. Further, the laser repetition rate was set to 10 Hz by a pulse picker (methods), ensuring the field $H_z$ sweeping sequence and the laser pulse sequence are synchronized (Fig. 1c).

One of the key results in this experiment is shown in Fig. 1d-e, where we compare the $M_z$ changes induced by laser with/without an in-plane magnetic field $H_x$. While a laser fluence around 1 mJ/cm$^2$ is applied, little differences can be found between the loops shown in Fig. 1b and d. Surprisingly, when we repeat sweeping $R$-$H_z$ loop with an additional small in-plane magnetic field, magnetization transition occurs stochastically from up to down and vice versa following each pulse, even resist the $H_z$ field direction (Fig. 1e). For more measurements of $R_{AHE}$ by sweeping $H_z$ at various laser fluences and $H_x$, see supplementary information 1. We further exclude the reliance on polarities of a tiny read current by inversing current flow directions and helicities of the light by rotating the laser polarizations. These results clearly indicate the significant modification of $M_z$ by joint effect of magnetic field $H_x$ and single-shot laser pulses. However, the modification of magnetization needs to be more stable, controllable therefore switchable.

**Single-pulse optical toggle switching of ferromagnetic films and hallbars**

To experimentally achieve toggle switching of magnetization, a systematic study was carried out. By adjusting laser power and balancing field $H_x$ and $H_z$ carefully, we achieved robust toggle switching events. Moreover, positive and negative sweeps of $H_x$ could identify the behavior is independent of the external magnetic field directions (supplementary information 2), which is remarkably distinct from that demonstrated in well-known spin-polarized current-induced manipulation[25–27]. Following an extensive investigation of perpendicularly magnetized samples with a wide range of bottom buffer layers X and varied thickness of CoFeB, we identified W (3)/CoFeB (0.8)/MgO (1.6)/Ta (1.6) as exhibiting the best energy-efficiency. Accordingly, this stack was implemented in further experiments and served as the free layer for our MTJ structures (Fig. 4a). Using the setup and sample mentioned above, we obtained the 100% toggle switching events shown in Fig. 2a, which is a typical segment from Fig. 2c. It shows the electrically detected state diagram of laser fluence (*x*-axis) and in-plane field (*y*-axis), where red areas present deterministic toggle switching behaviors and blue areas mean that the magnetization is partially switched.

The threshold fluences or minimum in-plane fields necessary to trigger magnetic reversal are determined by sweeping laser power or external fields (method). At each point, we recorded $R_{AHE}$ values lasting more than 20 seconds (100 laser shots). Thus, switching possibilities $P_{sw}$ (0%~100%) based on resistance variations were calculated as the color bar (supplementary information 3). Notably, in the blue region, the switching exhibits a stochastic character that may indicate a partial switching mode constrained by the Hallbar dimensions, a hypothesis that could be further corroborated by Kerr microscopy.

Figure 2b illustrates the Kerr images after each laser pulse, where the observed bullseye ring structures provide strong evidence for the precessional switching regime. For larger spot size and detailed power sweep, see extended data Fig. 1. Then we obtain another state diagram from Kerr microscopy shown in Fig. 2d, where the red region represents that the spot center area is toggled, while the blue region means not but the rings outside of center is toggled. Typical periodic behaviors can be observed both electrically and optically in the diagrams shown in Fig. 2c and 2d. However, notable differences exist, such as $P_{sw}$, switching thresholds, and their dependencies. We further conclude that the lateral dimensions of the crossbar structure can influence the magnetic domain distribution and, consequently, the switching ratio of $M_z$, as witnessed in supplementary information 4. Although an external magnetic field of at least 250 Oe is required to achieve toggle switching, this result suggests that the relatively small in-plane field could potentially be replaced by alternative methods.

**Transient thermal torque for coherent magnetization switching**

Up to now, these phenomena undoubtedly find some clues from a previous work[23] on bismuth-substituted yttrium iron garnet (Bi:YIG) ferrimagnetic films. However, so far there has been no experimental agreement in any ferromagnets. Here, we demonstrated the field-free magnetization switching of Ta/CoFeB (wedge)/MgO in nanosized devices by introducing a tilted anisotropy field (extended data Fig. 2). Although 100% toggle switching has not yet been achieved, this interesting form of all-optical switching in ferromagnets could stimulate advances in magnetic material engineering. Above all, we demonstrated that single-pulse optical switching can be applied to a practical material platform CoFeB/MgO, which is rare-earth-free, stable and unique to the MRAM industry. These results may also provide insights into the efficiency of current-induced magnetization switching at picosecond timescales[28], where ultrafast Joule heating is likely to play a critical role.

Next, we try to quantify the precessional mechanism by analytical solutions and ultrafast dynamics. By analogy to voltage-controlled magnetic anisotropy (VCMA)[13], the laser-induced modification, i.e. the ultrafast heating effect can be introduced as a transient change[23,24,27,29] in interfacial PMA. As illustrated in Fig. 3a, time- and temperature-dependent anisotropy plays a role as a modulation of the effective magnetic anisotropy field, $\Delta H_{k,eff}$, thereby triggering ultrafast magnetization dynamics. The effective field is derived from magnetic anisotropy energies and Zeeman energy, which are temperature dependent. The resulting thermal anisotropy torque is given by $\boldsymbol{\tau}_{\text{TAT}} = -\gamma \boldsymbol{M} \times \Delta \boldsymbol{H}_{\text{k,eff}}$, where $\gamma$ is gyromagnetic ratio. Inspired by the concept of the VCMA coefficient[13], we introduce the temperature derivative of the normalized magnetic anisotropy field, given by $\eta = d[H_{\text{k,eff}}(T)/H_{\text{k,eff}}(300\text{K})]/dT$, enabling us to reformulate the temporal evolution of magnetization $M(t)$ as a response to the transient thermal excitation $T(t)$:

$$\frac{d\boldsymbol{M}}{dt} = -\gamma \boldsymbol{M} \times [\boldsymbol{H}_{\text{ext}} + \boldsymbol{H}_{\text{k,eff}}(300\text{K})] + \alpha \boldsymbol{M} \times \frac{d\boldsymbol{M}}{dt} + \boldsymbol{\tau}_{\text{TAT}} \quad (1)$$

$$\boldsymbol{\tau}_{\text{TAT}} = \eta \cdot \Delta T \cdot H_{\text{k,eff}}(300\text{K}) \cdot \gamma \boldsymbol{M} \times \boldsymbol{m}_{\text{z}} \quad (2)$$

Notably, we treat $\eta$ as a constant, determined from the linear fitting of $H_{\text{k,eff}}(T)$ in 300-500 K range, based on our experiments and investigations (Fig. 3b and supplementary information 5, Table.1). This negative parameter can be directly extracted from the temperature dependence of $H_{\text{k,eff}}(T)$, providing a quantification of thermal control over the anisotropy field. For example, $\eta$ could be as large as $-0.01 \text{ K}^{-1}$ in a few atomic layers of FeCo with ideal MgO interfaces[30], which means a temperature rise $\Delta T$ of 50 K can reduce the magnetic anisotropy field by 50%. Here, the temperature profile $T(t)$ serving as a stimulus, was obtained from a numerical solution of the one-temperature (1T) model. According to the physical picture depicted in Fig. 3a, the effective magnetic anisotropy field $H_{\text{k,eff}}$ rapidly decreases to its minimum, and $\boldsymbol{\tau}_{\text{TAT}}$ reaches its peak at the very beginning and favors magnetization reversal. Therefore, precesisonal dynamics can be effectively induced by a sharp thermal jolt, but not by a slow and massive warming. In other words, $\boldsymbol{\tau}_{\text{TAT}}$ is not an integrative work accumulated over time, but rather a prompt, non-conservative torque that emerges from a rapid change in interfacial magnetic anisotropy. Our results indicate that the instantaneous lattice temperature rise is moderate ($\Delta T \sim 50$ K), yet sufficient to trigger significant precessional dynamics, ultimately enabling switching. Further, the evolution of $T(t)$ is, in turn, strongly governed by optical absorption and the thermal conductivity at the interface between the buffer layers and the substrate. The magnitude and rate of this thermal modulation are critical for determining the amplitude and timescale of precessions driven by $\boldsymbol{\tau}_{\text{TAT}}$, thereby dictating whether the magnetization can cross the equatorial plane within the first quarter of the precessional cycle (i.e., switched or unswitched). Based on the LLG and 1T models described in method and supplementary information 6-8, we performed macro-spin simulations that reproduce the magnetization reversal state diagram and consistently capture the striking features of precessional switching.

To probe the speed of magnetic dynamics, we traced magnetization $M_{\text{z}}$ over time using the time-resolved magneto-optical Kerr effect measurement setup (TR-MOKE, see methods.). External magnetic fields are applied nearly perpendicular to $\boldsymbol{m}_{\text{z}}$, existing tiny out-of-plane components allowing the switched magnetization to recover between every two laser pulses, which are required by the sampling scheme of pump-probe technique. As shown in Fig. 3c, we measured the hysteresis loop with static MOKE before the pump and the transient loop with TR-MOKE at 8.3 ps after pump respectively. At that moment, the magnetic anisotropy field $H_{k,eff}$ is significantly manipulated. Due to the incomplete closure of the transient hysteresis loops measured under varying in-plane magnetic fields, making it difficult to provide the complete temporal evolution of $H_{k,eff}(t)$ here. Nevertheless, this could be consistent with our simulation results (supplementary information 7), which predict the evolution of both $H_{k,eff}(t)$ and $M_s(t)$ over time. In both experiments and simulations, **M** is demagnetized first and then oscillates and decays in hundreds of picoseconds under ultrashort optical exposure. Further, as the external magnetic field increases, the amplitude of precession initially grows and is subsequently suppressed at higher fields (Fig. 3d). The phenomena clearly support the presence of a thermal anisotropy torque and agree well with the physical picture shown in Fig. 3a and the simulated results according to equation (1)-(2). We would emphasize that small variations of damping and strain may influence a little on the magnetic dynamics. To date, our model incorporates only the thermal effect on interfacial magnetic anisotropy. However, the possibility of light-controlled interfacial magnetic anisotropy due to photon-phonon interactions and magneto-acoustic coupling cannot be ruled out yet.

**Demonstrating opto-spin integration through magnetic tunnel junctions**

The observed ultrafast toggle switching in single ferromagnetic CoFeB layers under single laser pulses, driven by thermal anisotropy torque $\boldsymbol{\tau}_{TAT}$ may advance ultrafast opto-spintronics. Motivated by this mechanism, we sought to implement it in practical magnetic tunnel junctions (MTJs). To this end, we fabricated top-pinned MTJs on a transparent glass substrate (methods), providing convenient optical access to the free layer from backside of the transparent substrate, as illustrated in Fig. 4a. Unlike the simplest W/CoFeB/MgO structure with strong interfacial PMA, the fabricated stack incorporates a reference layer and multilayered synthetic antiferromagnetic structures, as depicted in Fig. 4c. Figure 4b displays the *R-H* measurement, revealing a tunneling magnetoresistance (TMR) exceeding 110%, a record for optically switchable MTJs. Additionally, we reproduced laser-induced toggle switching in these complete MTJ stacks using Kerr microscopy (Fig. 4d), although the contrast is relatively low due to imaging through the transparent substrate and the complexity of multilayered magnetic system. A more likely reason is that under an in-plane magnetic field, the magnetization has already tilted, reducing the intensity of the Kerr signal. Alternatively, the switching ratio might not be sufficiently high. Thus, we fabricated micrometer-scale MTJ devices to validate our assumption via TMR measurements. Indeed, we observed that the TMR decreased by 75% with 4-um-diameter MTJ under a 2.1 kOe in-plane magnetic field (Fig. 2b). In Fig. 4e, we show a time trace of reversible laser-induced magnetization switching (~2 Hz), with a toggling ratio of 52.6%. The partial switching behavior might be attributed to the influence of the possible stray field from the reference layer or angular momentum transfer induced by ultrafast hot electrons [21]. However, a MgO thickness of 1.6 nm may stop most of the spin injection carried by laser-excited hot electrons[20], and optical manipulation is not observed at this fluence in the absence of in-plane magnetic field.

Figure 4f summarizes the switching threshold fluence for W/CoFeB/MgO devices as a function of pulse durations under various in-plane magnetic fields. The measured laser energy fluence threshold can reach as low as 0.22 mJ/cm$^2$, corresponding to an estimated energy consumption of 17 fJ/bit for a 100-nm-diameter dot (Fig. 4f right y axis). The ultra-low energy consumption highlights the potential of thermal anisotropy torque for energy-efficient memory and logic applications. The experimental observations align well with the thermal anisotropy torque $\tau_{\text{TAT}}$ regime. In experiments, we have achieved single-shot optical switching for pulse durations ranging from 50 fs to 16.29 ps, with increasing laser fluence thresholds. Shorter laser pulses can more efficiently heat the electron bath, leading to a more dramatic lattice temperature rise $\Delta T$ during the heating phase (supplementary information 6). Our simulation further indicates that this torque $\tau_{\text{TAT}}$ remains effective for pulse widths up to approximately 150 ps (supplementary information 8). Above all, this principle has been successfully implemented in rare-earth-free MTJs, compatible with laser systems with pulse widths ranging from tens of femtoseconds to hundreds of picoseconds, opening wider opportunities for opto-magnetic integrated applications.

**Conclusions**

The ability to reverse a ferromagnetic layer on picosecond timescales using a single laser pulse paves the way to a new generation of highly stable photonic memory and storage technologies. Our prototype's layer and device structures are simple, scalable, and highly compatible with existing memory industry practices. Such ultralow energy consumption places thermal anisotropy torque $\tau_{\text{TAT}}$ among one of the most competitive writing technologies compared with conventional hard disk drives (HDD) and MRAMs. We note that, because this ultrafast heating comes from the light illuminating uniformly on the thin-film surface, further engineering using plasmonic focusing enhancements may significantly improve energy efficiency. Moreover, by substituting the external magnetic field with a tilted internal anisotropy field, we provide a clear route towards realizing full all-optical switching in a single layer ferromagnet, which may address one of the central challenges in optical switching technologies. Our results thus bridge the gap between spintronics and ultrafast magnetism. Additionally, our approach demonstrates that laser pulses can efficiently access the magnetic free layer through transparent substrates, facilitating single-pulse optical writing. This capability allows chip- or wafer-level bonding, opening avenues for novel architectures featuring optical switching and electrical detection in magnetic domain or combining optical and electrical control of magnetism. In this way, arrays of photonic-spintronic memory devices could be addressed via integrated optical waveguides, achieving possibly ultimate speed and minimal energy dissipation while maintaining high read-out sensitivity.

## Acknowledgements

We acknowledge the nanofabrication facility in Beihang Nano for support with device fabrication and characterization. We thank Dr. Zongxia Guo from Laboratoire Albert Fert, Université Paris-Saclay, Dr. Yong Xu, Mr. Xiaoyang Chang and Mr. Ziyue Wang from Beihang University, Dr. Honglei Du and Mr. Shaoxin Li from Truth Instruments Corp., Ms. Cong Zhang, Ms. Wenwen Wang, Mr. Zongwei Guo, Mr. Xiaodong Jiang and Mr. Yinchao Qin from Truth Memory Corp., for their helpful discussions, assistance in sample fabrication and set-up engineering. This work is supported by the National Key Research and Development Program of China (No. 2022YFA1402604), the National Natural Science Foundation of China (No. 12104031 and T2394473), the Fundamental Research Funds for the Central Universities (No. YWF-23-Q-1089) and Support Plan for High-Level Student Science and Technology Innovation Teams from Beihang University (No. 501XSKC2024142001 and 501XSKC2025142001).

## Author contributions

Weisheng Z. initialized, conceived and supervised the study. Chen X., Boyu Z., Xiangyu Z. and Weisheng Z. planned the experiments. Yuxuan Y., Chen X., Yu H. and Yan H. designed and fabricated the samples with the help of Shiyang L., Kaihua C. and

Hongxi L. Chen X., Boyu Z., Xiangyu Z., Jiaqi W. and Dinghao M. designed and built the set-up with the help of Xueying Z. and Weisheng Z. Chen X., Boyu Z., Xiangyu Z. Yuxuan Y., Jiaqi W. Wenlong C. and Dinghao M. performed the magnetic microscopy and optical switching measurements with suggestions from Bert K., Youguang Z. and Weisheng Z. Chen X. and Yuting G. performed time-resolved measurements with help of Xiangyu Z. and Xianyang L. Chen X., Boyu Z. and Rui X. carried out macrospin and thermal modeling with help of Daoqian Z. and Pierre V. Chen X., Boyu Z., Xiangyu Z. Yuxuan Y., and Weisheng Z. wrote the manuscripts. All the authors discussed the data analysis and commented on the manuscript.

**Competing interests**

The authors declare no competing interests.

**Additional information**

All study data are included in the article and supplementary information. Additional materials, raw data and code can be available from the corresponding author upon request.

**Displayed items**

Figure 1 | Pulsed laser-induced manipulation of perpendicular magnetization in CoFeB/MgO Hallbar devices.

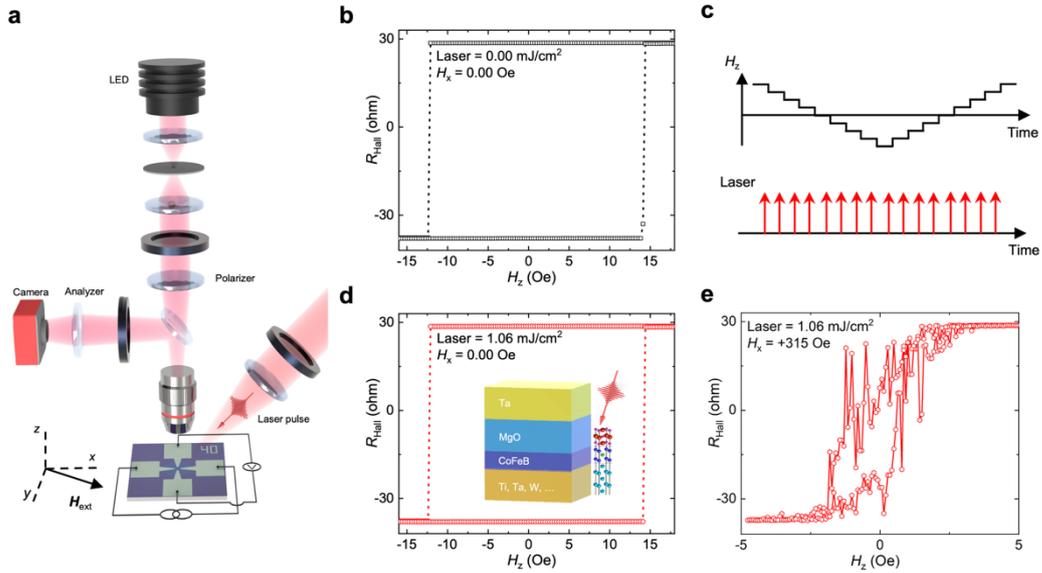

**Figure 1 | Pulsed laser-induced manipulation of perpendicular magnetization in CoFeB/MgO Hallbar devices. a**, Schematic of the measurement set-up. A train of laser pulses were introduced to the devices under test using pulse widths of 50 fs~16.29 ps and repetition rates of 1 Hz~1 kHz. The magnetization changes can be probed by both Kerr microscopy and AHE resistance. **b**, Static AHE resistance ($R_{Hall}$) vs. out-of-plane magnetic field ($H_z$) loop measured without laser excitation. **c**, Schematic of the laser pulse sequence (10 Hz) and out-of-plane magnetic field ($H_z$) sweep (100 ms per step). Electrical measurements were conducted after each step. **d**, $R_{Hall}$ - $H_z$ loop under laser excitations as described in **c**. Inset is the schematic structure of the multilayered X/CoFeB/MgO/Ta thin films. **e**, $R_{Hall}$ - $H_z$ loop under laser excitations with an additional small in-plane magnetic field $H_x$, which is distinct from the squared loops shown in **b** and **d**.

Figure 2 | Laser-induced precessional switching in ferromagnetic thin films and Hallbar devices.

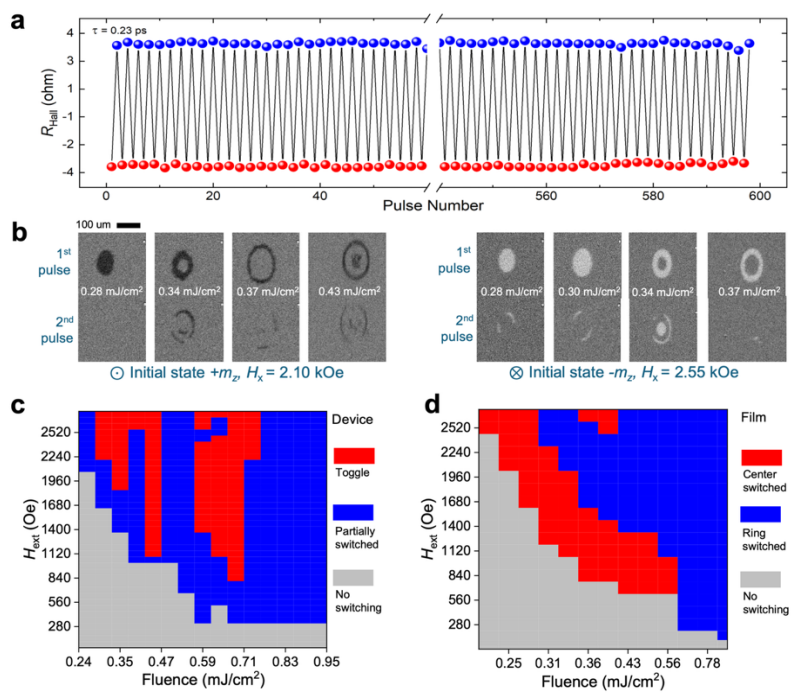

**Figure 2 | Laser-induced precessional switching in ferromagnetic thin films and Hallbar devices. a**, Counting measured anomalous Hall resistance ($R_{Hall}$) under 5-Hz pulsed laser excitations, exhibiting perfect toggle switching behavior with nearly 100% optical switching ratio compared to that switching by magnetic field. **b**, Kerr microscopy images (initial state, up and down respectively) show the magnetic domain structure switching after 1st and 2nd laser shot, under increased laser fluences and different in-plane magnetic fields ($H_x$ = 2550 Oe or 2100 Oe), revealing the striking bullseye ring patterns, indicating precessional switching mechanism. **c**, Magnetization switching phase diagram of thin film as a function of laser fluence and in-plane magnetic field, identifying different switching regions, marked as "Bistable toggle"-red, "Stochastic partially switched"-blue and "No switching"-grey respectively. Each point value in **c** is obtained from observations in **a**. **d**, Magnetization switching phase diagram of thin film as a function of laser fluence and in-plane magnetic field, identifying different switching regions, marked as "Center toggle switching"-red, "Outer Rings"-blue and "No switching"-grey respectively. Each point value in **d** is obtained from observations in **b**.

Figure 3 | Thermal anisotropy torque activated by ultrafast heating.

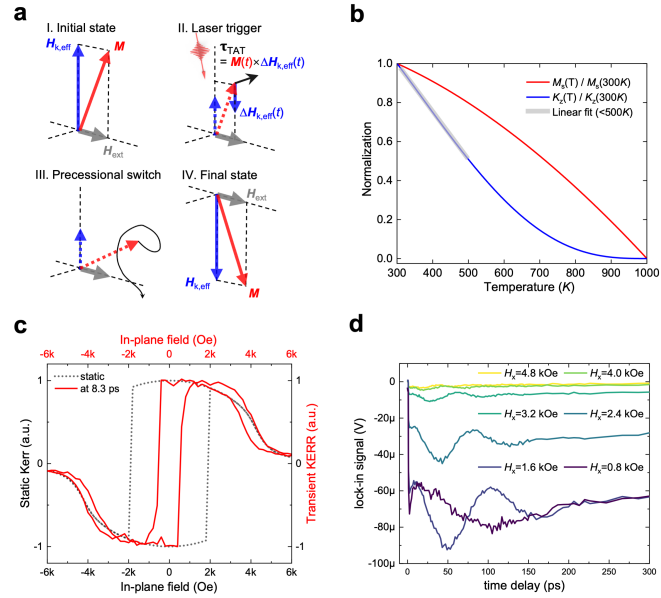

**Figure 3 | Thermal anisotropy torque activated by ultrafast heating. a**, Illustration of the magnetization vector **M** dynamics under the influence of the thermal anisotropy torque $\tau_{\text{TAT}}$, the effective magnetic anisotropy field $H_{k,\text{eff}}$, and the external magnetic field $H_{\text{ext}}$. The laser reduces the $H_{k,\text{eff}}$ and $M_s$ at the beginning, a remarkable torque is resulted from $M \times \Delta H_{k,\text{eff}}$, then driving precessions. The magnetization vector should tilt towards the in-plane direction with a large precession angle and may cross the equational plane before the lattice cools down to the room temperature. **b**, Normalized magnetization $M_s$ and PMA energy density $K_z$ as a function of temperature $T$. The $K_z$ is supposed to be much more sensitive to temperature variations than $M_s$ in ultrathin FeCo magnetic layers. The grey line gives a linear quantification of d$K_z$/d$T$ during 300 K ~ 500 K. **c**, Normalized static $M$-$H_x$ loop and transient $M$-$H_x$ loop at a pump-probe delay of 8.3 ps. A reduction of $H_{k,\text{eff}}$ can be observed. **d**, Time-resolved MOKE measurements of unswitched magnetic dynamics for various in-plane magnetic fields.

Figure 4 | Single shot optical magnetization switching in ferromagnetic MTJs.

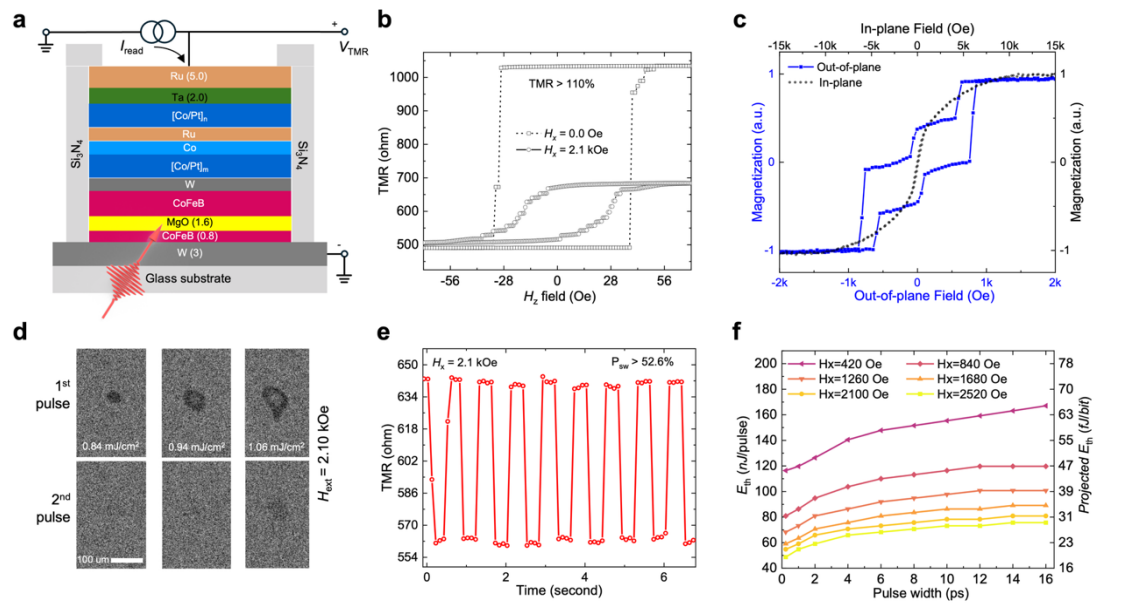

**Figure 4 | Single shot optical magnetization switching in ferromagnetic MTJs. a**, Cross-sectional schematic of the top-pinned MTJ full stack deposited on a glass substrate. The double-polished transparent substrate provides an optical access to the magnetic free layer W/CoFeB/MgO. The sample was flipped under test using the set-up mentioned in Fig. 1a. We use wire bonding to measure the tunneling magnetoresistance (TMR) at frontside. **b**, Static tunnel *R-H* loop showing a TMR ratio more than 110% at zero in-plane magnetic field (square) and around 27% at 2.1 kOe in-plane magnetic field (circle). **c**, Normalized *M-H* loops under out-of-plane magnetic field (blue square, solid line) and in-plane magnetic field (red dashed line) using vibrating sample magnetometer (VSM). **d**, Kerr microscopy observed from the back side of sample (via substrates), showing the toggle switching behavior after 1st and 2nd laser shot, under increased laser fluences and different in-plane magnetic fields ($H_x$ = 2550 Oe or 2100 Oe), reproducing phenomena shown in Fig. 2b. **e**, Demonstration of toggle switching of magnetization with subsequent laser pulses, achieving a switching ratio more than 52.6%. **f**, Dependence of the energy thresholds for switching the magnetic layer W/CoFeB/MgO on the laser pulse widths ranging from 0.23 ps to 16.29 ps. The laser pulse energy consumption for switching is calculated (right *y*-axis) assuming the device is projected to a 100-nm-diameter magnetic dot.

## Online items

Extended data Figure 1 | Kerr microscopy images of single-pulse switching at various pulse durations.

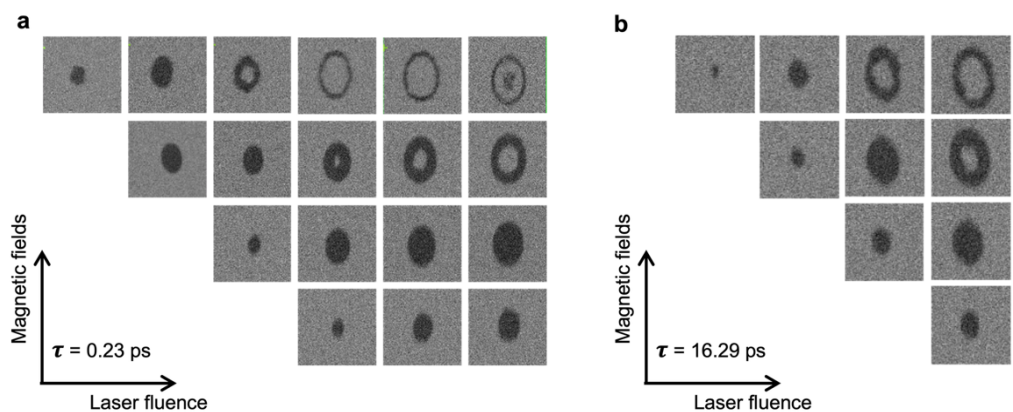

**Extended data Figure 1 | Kerr microscopy images of single-pulse switching at various pulse durations**. **a**, Domain patterns as a function of in-plane magnetic field and laser fluence for a 0.23 ps pulse duration. **b**, Corresponding results for a 16.29 ps pulse duration. All measurements were performed on W (3)/CoFeB (0.8)/MgO (1.6)/Ta (1.6) thin films.

Extended data Figure 2 | Field-free optical switching (All-optical switching, AOS) of Ta/CoFeB(wedge)/MgO nanodot devices.

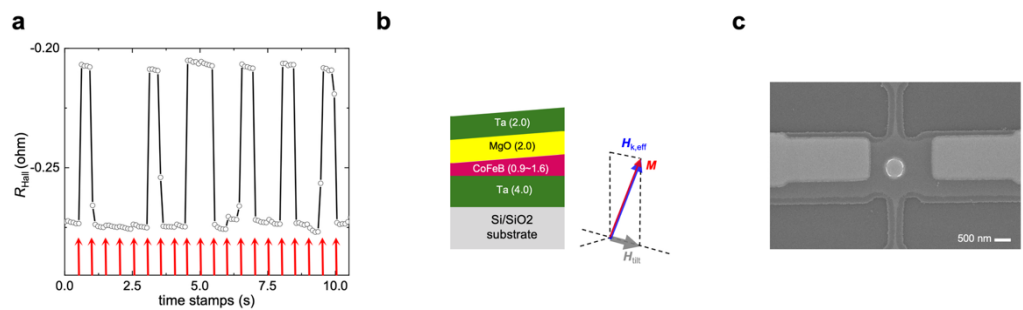

**Extended data Figure 2 | Field-free optical switching (All-optical switching, AOS) of Ta/CoFeB(wedge)/MgO nanodot devices. a**, Time-evolution of $R_{Hall}$ under a train of laser pulses (2 Hz), with a fixed laser fluence of 0.9 mJ/cm$^2$ and zero in-plane magnetic field. It provides approaches (in-plane effective magnetic field from material engineering) to achieve all-optical switching of a single layer ferromagnet down to nanoscales. **b**, Schematic of side view of the film stack (numbers in nanometer), Ta (4)/CoFeB (wedged,0.9-1.6)/MgO (2)/Ta (2), resulting a tilted magnetic anisotropy field. The measured position for **a** is around 1.3 nm. **c**, Top view of the nanodot devices in a crossbar structure, a scanning electron microphotograph. The white scale bar is 500 nm.

# Methods

## Sample preparation

For the initial Hall effect measurements, a series of thin films with the structure X/ $Co_{20}Fe_{60}B_{20}$ (0.8~1.4)/MgO (2)/Ta (2) (thicknesses number in nanometers) were deposited on thermally oxidized Si (001) substrates using a sputtering system with a base pressure of 5e-7 Torr. Here, X represents different buffer layers: Ta (5), Ti (5), W (3), and a Ta (2)/Ti (3) bilayer. The deposition rates for each material were 3.75 min/nm for CoFeB and 5.95 min/nm for MgO at an Ar pressure of 1e-3 Torr.

The MTJ stack with the structure：Ta (0.15)/W (3)/ $Co_{20}Fe_{60}B_{20}$ (0.8)/MgO (1.6)/ $Co_{20}Fe_{60}B_{20}$ reference layer/W/[Co/Pt]$_m$/Co/Ru/[Co/Pt]$_n$/Ta (2)/Ru (5) (thicknesses number in nanometers) was deposited on double side polished glass substrates (BK7) with sizes of 2 inches using a Singulus ROTARIS 200-mm magnetron sputtering system at a base pressure of 3e-7 Torr. The MgO barrier was deposited by RF sputtering. The entire stack was insitu annealed at 350 °C for 20 min in the vacuum chamber without applying magnetic field.

## Device fabrication

The Hall bar devices were patterned using two steps of UV lithography combined with Ion Beam Etching (IBE) and sputtering. The devices had lateral dimensions of 5 *um*, 10 um and 20 um. We used 5-um-width crossbars for most measurements shown in Fig. 1 and 2, while 20-um-width devices were used for observing Kerr imagines. After the 2$^{nd}$ negative lithography, the electrodes with Ti (10)/Au (70) were sputtered on the samples and followed by a lift-off process. For nanodot devices shown in Extended data Fig. 2, additional steps were needed: nanosize pattern by electron beam lithography (EBL), etching and stopping after the MgO by IBE and insulating layer deposition by chemical vapor deposition (CVD).

Microsized MTJ devices (pillars) with diameters of 3 um, 4 um, and 5 um were patterned using three steps of UV lithography combined with IBE. We used 4-um MTJs for electrical measurements shown in Fig. 4b and 4e. The bottom electrodes were patterned first, followed by the MTJ pillars. The remaining bottom electrode thickness after the MTJ pillar etching was approximately 3 nm. Then, the devices were then covered with a 60 nm thick SiO2 insulating layer deposited by CVD and followed by a lift-off process. Finally, after final negative lithography, a Ti (50)/AlCu (150)/TiN (10) top electrode was deposited by a DC sputtering system and patterned using a standard lift-off process. The transparent substrate was estimated to allow more than 80% transmission into the sample when the coming light is perpendicular to its surface.

## Magneto-optic Kerr microscope for optical and electrical measurements

With engineering support from Truth Instruments Co., Ltd., Qingdao, China, we co-developed a probe station for optical writing and electrical reading based on a magneto-optic Kerr microscope (MOKE) system with vector magnetic fields. This multifunctional probe station allows us to check the magnetic domains distribution when monitoring the electrical characteristics. The Hall resistances of the

X/CoFeB/MgO/Ta samples was measured using a typical 4-probe Hall effect measurement setup with a 30 μA reading currents applied along *x*-direction. Then the Hall voltages at two terminals across *y*-direction were then measured. External magnetic fields were applied using electromagnets in both out-of-plane and in-plane directions. A small out-of-plane field was used in combination with the in-plane field to ensure proper balancing. The static magnetic properties of the MTJ devices, including the TMR, were characterized using a four-point probe method with the magnetic field applied out-of-plane (±*z*-direction). The TMR was calculated as TMR = ($R_{AP}$–$R_P$)/$R_P$, where $R_{AP}$ and $R_P$ are the resistances in the antiparallel and parallel magnetization states, respectively.

For ultrafast excitations, two different laser systems were integrated into the set-up. We used a Ti:Sapphire laser system (center wavelength 800 nm) generating pulses with durations about 50 fs at a repetition rate of 0~1 kHz. The laser pulse repetition rates are altered and picked by an arbitrary function generator (AFG). For longer pulse durations, an Er-doped fiber laser system (center wavelength 1030 nm), generating pulses with durations ranging from 180 fs to 16.29 ps, with a repetition rate of 0~100 kHz. The laser beam was focused onto the sample surface using a lens (focus length = 150 mm) with a spot size of 300 μm. The laser fluence was controlled using a motorized half-wave plate before a linear polarizer. The polarization of the pump laser was set to linear or circular by quarter-wave plate. The change of magnetization states of the Hall bar and MTJ devices were investigated by monitoring the sudden change in the AHE and TMR signal upon laser excitation, respectively. The change in resistance can be validated by Kerr microscopy at the same time. The whole measurement was performed this multifunctional probe station.

**TR-MOKE Measurements**
Time-resolved magneto-optic Kerr microscopy (TR-MOKE) measurements were performed using a typical pump-probe configuration at a repetition rate of 1 kHz. In the measurements, the sample was exposed by a pump pulse with a center wavelength 800nm, a duration of 50 fs and a spot size of typically 450 μm. Meanwhile, a probe pulse, which arrived at the sample after a time delay with a spot size of 150 um. In-plane magnetic fields were applied during the measurement with small perpendicular component allowing to reset the perpendicular magnetization between two pump pulses. For transient loop measurements, in-plane magnetic fields were swept at a certain pump-probe delay. The TR-MOKE measurement and data processing method is typical and consistent with other studies.

**Macromagnetic Simulation**
We develop a thermal-modified LLG equations (1) and (2) mentioned in the main text, an analogy to the voltage-gated case. Equations (1) and (2) are derived from:

$$\boldsymbol{H}_{\text{eff}}(T) = \boldsymbol{H}_{\text{k,eff}}(T) + \boldsymbol{H}_{\text{ext}} = \left[\frac{2K_z(T)}{\mu_0 M_S(T)} - M_S(T)\right]\boldsymbol{m}_z + \boldsymbol{H}_{\text{ext}} \qquad (3)$$

$$\frac{d\boldsymbol{M}}{dt} = -\gamma \boldsymbol{M} \times \boldsymbol{H}_{\text{eff}}(T) + \alpha \boldsymbol{M} \times \frac{d\boldsymbol{M}}{dt} \tag{4}$$

$$\boldsymbol{H}_{\text{k,eff}}(t) = \boldsymbol{H}_{\text{k},eff}(300\text{K}) \cdot [1 + \eta \cdot T(t)] \tag{5}$$

Where $K_z(T)$ is PMA energy density and $M_S(T)$ is saturation magnetization. They are temperature-dependent, and their derivatives with respect to temperature are usually inconsistent. The key point is that equation (2) introduces a thermal anisotropy torque coefficient $\eta$, which can be experimentally determined in different magnetic material structures, like the efficiency of voltage control on magnetic anisotropy (VCMA) and the spin Hall angle. First, we use 1-T model to simulate the temperature as a function of time $T(t)$. Then the temperature pulse was served as a stimulus in modified LLG equation, where the temporal profile of effective magnetic anisotropy field $H_{\text{k,eff}}(t)$ is expressed as equation (5). All the simulations were conducted by Python (supplementary information 6-8).